\documentclass[conference]{IEEEtran}
\IEEEoverridecommandlockouts
\usepackage{amsmath,amssymb,amsfonts}
\usepackage[pdftex]{graphicx}
\usepackage{algorithmic}
\usepackage{multirow}
\usepackage{graphicx}
\usepackage{threeparttable}
\usepackage[table,xcdraw]{xcolor}
\usepackage{xcolor}
\usepackage{xspace}
\usepackage{enumitem}
\usepackage{acronym}
\usepackage[binary-units=true,per-mode=symbol]{siunitx}
\usepackage[a4paper, total={184mm,239mm}]{geometry}
\usepackage[shortcuts]{extdash}
\usepackage{fixfoot}
\usepackage[hyphens]{url}
\usepackage[hidelinks]{hyperref}
\usepackage{svg}
\usepackage{textcomp}
\usepackage{xcolor}
\usepackage{tabularx,booktabs}
\usepackage{flushend}
\usepackage{balance}
\usepackage{textcomp}

\usepackage{tabularx,booktabs}
\usepackage{diagbox}
\newcolumntype{C}{>{\centering\arraybackslash}X}
\def\BibTeX{{\rm B\kern-.05em{\sc i\kern-.025em b}\kern-.08em
    T\kern-.1667em\lower.7ex\hbox{E}\kern-.125emX}}

\renewcommand{\baselinestretch}{0.85}

\begin{document}

\title{RedMulE: A Compact FP16 Matrix-Multiplication Accelerator for Adaptive Deep Learning on RISC-V-Based Ultra-Low-Power SoCs\\
}
\acrodef{FP}{floating-point}
\acrodef{DL}{Deep Learning}
\acrodef{ML}{Machine Learning}
\acrodef{NN}{neural network}
\acrodef{BNN}{Binary Neural Network}
\acrodef{CNN}{Convolutional Neural Network}
\acrodef{DNN}{Deep Neural Network}
\acrodef{RNN}{Recursive Neural Network}
\acrodef{QNN}{Quantized Neural Network}
\acrodef{MAC}{Multiply-Accumulate}
\acrodef{HWPE}{Hardware Processing Engine}
\acrodef{PE}{Processing Element}
\acrodef{FMA}{Fused Multiply-Add}
\acrodef{SoA}{State-of-the-Art}
\acrodef{TCDM}{Tightly-Coupled Data Memory}
\acrodef{LIC}{Logarithmic Interconnect}
\acrodef{FSM}{Finite State Machine}
\acrodef{NNZ}{Non-Zero Elements}
\acrodef{AI}{Artificial Intelligence}
\acrodef{HCI}{Heterogeneous Cluster Interconnect}
\acrodef{IoT}{Internet of Things}
\acrodef{ISA}{Instruction Set Architecture}
\acrodef{DSP}{Digital Signal Processing}
\acrodef{DMA}{Direct Memory Access}
\acrodef{FSM}{Finite State Machine}
\acrodef{FIFO}{First-In First-Out}
\acrodef{GPU}{Graphic Processing Unit}
\acrodef{SoC}{System on-Chip}
\acrodef{CL}{Continual Learning}
\acrodef{PULP}{Parallel Ultra-Low-Power}
\acrodef{MCU}{Microcontroller Unit}

\newcommand{\todo}[1]{\noindent\textit{\color{red}\textbf{}~#1}}
\newcommand{\new}[1]{\noindent{\color{blue}\textbf{}~#1}}
\newcommand{\revise}[1]{\noindent{\color{orange}\textbf{}~#1}}

\newcommand{\riscv}{RISC\=/V\xspace}
\newcommand{\etal}{\textit{et al.}\xspace}
\newcommand{\figref}[1]{Fig.~\ref{#1}}
\newcommand{\tabref}[1]{Table~\ref{#1}}
\newcommand{\dbit}[1]{{#1}\=/\si{\bit}}
\newcommand{\gf}{\textsc{Glo\-bal\-found\-ries}~22FDX\xspace}



\author{\IEEEauthorblockN{Yvan Tortorella\IEEEauthorrefmark{1}, Luca Bertaccini\IEEEauthorrefmark{2}, Davide Rossi\IEEEauthorrefmark{1}, Luca Benini\IEEEauthorrefmark{1}\IEEEauthorrefmark{2}, Francesco Conti\IEEEauthorrefmark{1}}
\IEEEauthorblockA{\IEEEauthorrefmark{1}University of Bologna, Bologna, Italy. \textit{Email: \{yvan.tortorella, davide.rossi, luca.benini, f.conti\}@unibo.it}
\IEEEauthorblockA{\IEEEauthorrefmark{2}ETH Zurich, Zurich, Switzerland. \textit{Email: \{lbertaccini, lbenini\}@iis.ethz.ch}} 
}}

\maketitle

\begin{abstract}
The fast proliferation of extreme-edge applications using Deep Learning (DL) based algorithms required dedicated hardware to satisfy extreme-edge applications' latency, throughput, and precision requirements. While inference is achievable in practical cases, online finetuning and adaptation of general DL models are still highly challenging.
One of the key stumbling stones is the need for parallel floating-point operations, which are considered unaffordable on sub-100\,mW extreme-edge SoCs. We tackle this problem with RedMulE (Reduced-precision matrix Multiplication Engine), a parametric low-power hardware accelerator for FP16 matrix multiplications - the main kernel of DL training and inference - conceived for tight integration within a cluster of tiny RISC-V cores based on the PULP (Parallel Ultra-Low-Power) architecture. In $\mathbf{22}$\,nm technology, a 32-FMA RedMulE instance occupies just $\mathbf{0.07}$\,mm\textsuperscript{2} ($\mathbf{14}$\% of an 8-core \riscv cluster) and achieves up to $\mathbf{666}$\,MHz maximum operating frequency, for a throughput of $\mathbf{31.6}$\,MAC/cycle ($\mathbf{98.8}$\% utilization). We reach a cluster-level power consumption of $\mathbf{43.5}$\,mW and a full-cluster energy efficiency of $\mathbf{688}$\,16-bit\,GFLOPS/W. Overall, RedMulE features up to $\mathbf{4.65\times}$ higher energy efficiency and $\mathbf{22\times}$ speedup over SW execution on 8\,\riscv cores.
\end{abstract}


\section{Introduction \& Related Works}
In  the  last  few  years,  the  amount  of  \ac{IoT} devices connected and executing \ac{ML} and,  in  particular, \ac{DL}  based  algorithms  like \acp{DNN}  has  considerably  increased. Moving  the  computation  from  data centers  to  energy-efficient \ac{IoT}  end-nodes  helps  lower  the  amount  of  data  sent  over the  network,  improve  energy  efficiency,  and  prevent network congestion. Extreme-edge \ac{ML} and Tiny-\ac{ML} use efficient \ac{IoT} endpoints to move data processing from the cloud to the edge. 

Extreme-edge inference is achievable in practical cases since it can be performed with low precision integer operations that help increase the energy efficiency, reduce the memory footprint and the area overhead, with reduced accuracy loss~\cite{edge_intell, ml_survey}. Eyeriss~\cite{eyeriss} is a hardware accelerator designed for \acp{CNN} inference with INT16 arithmetic and implemented in \SI{65}{nm} technology. Zeng \etal \cite{rssa} also designed an accelerator for \acp{CNN} inference in the same technology using INT8, and the accelerator features $1.8\times$ higher power consumption, $14.5\times$ higher energy efficiency, and $25\times$ higher throughput than Eyeriss. Also, EIE~\cite{eie} is an inference chip based on INT8 arithmetic designed for compressed \acp{DNN} and implemented in \SI{45}{nm} technology, but characterized by \SI{590}{mW} of average power consumption. Simba~\cite{simba} is a \acp{DNN} inference design implemented in \SI{16}{nm} technology and featuring \SI{9.1}{TOPS/W} energy efficiency with INT8 arithmetic.

All the chips mentioned so far are examples of accelerators designed only for extreme low-power inference, while on-chip training is still challenging because it imposes restrictive data accuracy and precision requirements. Single-precision and double-precision \ac{FP} operations provide sufficient range and dynamic, but are highly time and energy-consuming. \acp{GPU} like the NVIDIA Ampere A100~\cite{a_100} and HPC-oriented designs like Manticore~\cite{manticore} are examples of high-performance platforms for \ac{DL} training providing industry-leading computing performances at the cost of high power consumption and area occupation. These facts limit their application on edge, imposing severe limitations on implementing novel learning algorithms~\cite{cont_learn} on extreme-edge end nodes. For this reason, a strong effort was recently made to adapt learning algorithms to low-precision \ac{FP}16~\cite{nvidia_mixp, intel_mixp}, and \ac{FP}8~\cite{hyb_fp8} with no accuracy loss, showing that massive gains can be achieved by lowering the precision to just the right amount needed~\cite{trans_fp}.

Cambricon-Q~\cite{cambricon} provides an example of a training-oriented chip for high accuracy and energy efficiency, but their training is based on \dbit{8} fixed-point arithmetic, while it is well-known that most common training algorithms require floating-point operations. IBM~\cite{ibm,rapid} proposes an AI computing platform for training with \ac{FP}16 and hybrid \ac{FP}8 operands at the cost of overall power consumption that is not affordable on sub-100 mW extreme-edge devices.
Anders~\etal~\cite{anders} propose a reconfigurable accelerator for matrix-multiplications supporting mixed-precision computations, targeting training-oriented applications thanks to low power consumption, low area occupation, and \ac{FP}16 arithmetic.

\begin{figure}[t]
\centerline{\includegraphics[scale=0.78]{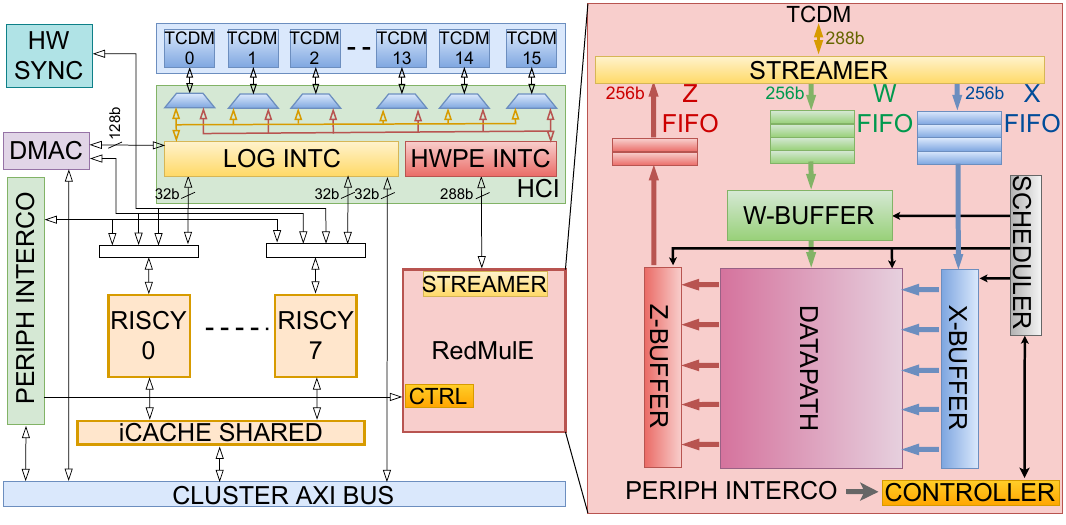}}
\caption{\ac{PULP} cluster overview with a focus on RedMulE architecture.\expandafter}
\label{fig:Pulp Cluster}
\end{figure}

\begin{figure*}[htbt]
\centerline{\includegraphics[width=\textwidth]{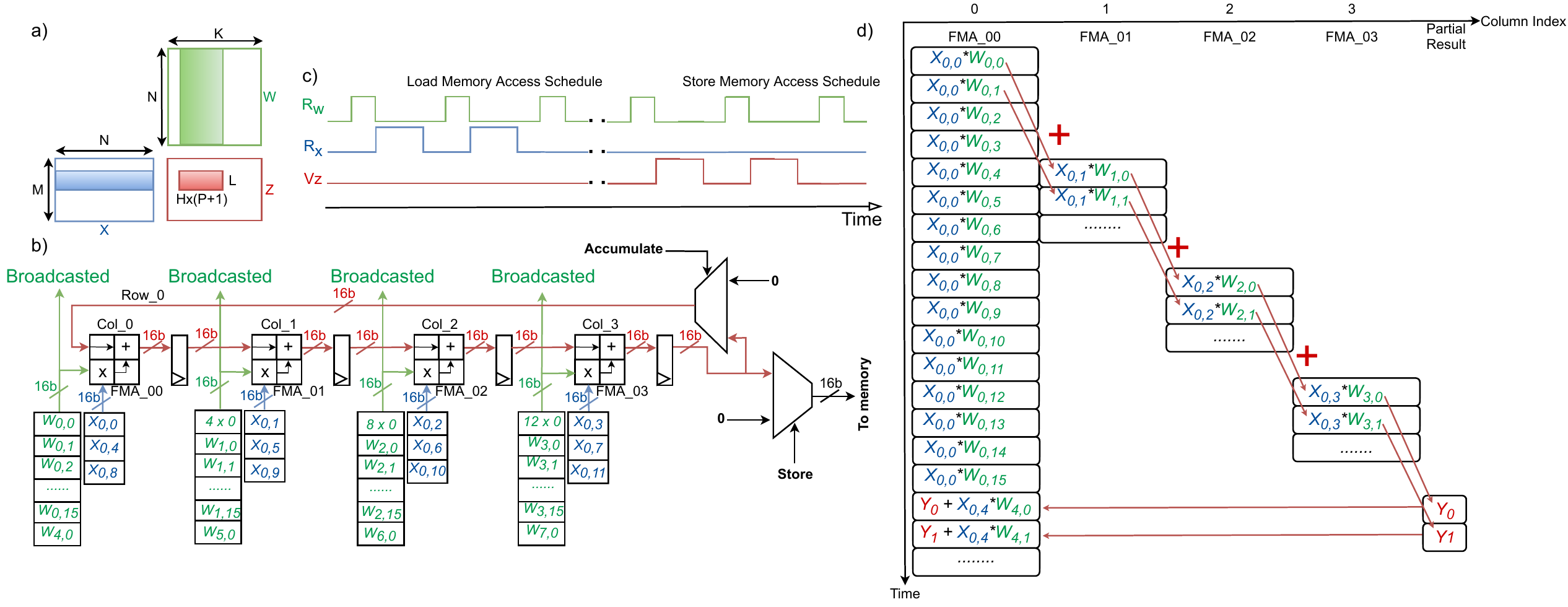}}
\caption{a) Matrix-multiplication execution; b) Row of \acp{FMA} within RedMulE Datapath; c) Memory access schedule in load/store mode described in terms of R (ready) and V (valid) handshake signals; d) Pipeline evolution within a row of \acp{FMA}.
}
\label{fig:architecture}
\end{figure*}
We present \textit{RedMulE} (Reduced-precision matrix Multiplication Engine), the first tightly-coupled and parametric hardware accelerator for \ac{FP}16 (IEEE binary16) matrix multiplication (the main kernel in online learning), designed to be integrated into a \ac{PULP}~\cite{pulp} cluster. It is optimized for better data reuse targeting ultra-low-power unified \ac{FP}16 training and inference on edge. We prototyped our design within an 8-core \ac{PULP} cluster in \SI{22}{nm}\,CMOS technology, targeting a RedMulE instance with 32 \acp{FMA}~\cite{fpnew}. We show that RedMulE occupies only \SI{0.07}{mm2} (14\% of the entire cluster), achieves up to $22\times$ speedup, and $4.65\times$ higher energy efficiency than a software counterpart running on the 8 RISC-V cluster cores.



\section{Architecture}
\subsection{PULP Cluster}
In Fig.~\ref{fig:Pulp Cluster}, we show the architecture of a PULP cluster featuring 8 \riscv cores and enhanced with a tightly-coupled accelerator called \ac{HWPE}~\cite{xnor}, that shares the system memory with the cluster cores and is software-programmed by the cores. The \ac{TCDM} is accessible via a single-cycle latency \ac{HCI} designed for synergistic operation of the \riscv cores and \acp{HWPE}. The HCI features two separated branches, \textit{logarithmic} and \textit{shallow}: the logarithmic branch allows all-to-all single-cycle access from \dbit{32} initiator ports (cores, DMA) to each of the word-interleaved memory banks; conflicts are managed by granting only one initiator per bank, with a round-robin scheme.
The shallow branch features a single 288-bit port, routed to 9 adjacent memory banks treated like a single 288-bit bank without arbitration.
The \ac{TCDM} banks are connected to the two branches via a set of multiplexers, which grant access to one branch or the other according to a configurable-latency starvation-free rotation scheme.

\subsection{RedMulE architecture}
RedMulE targets matrix multiplications of the kind:
\begin{equation*}
\mathbf{Z}=\mathbf{X}\cdot\mathbf{W}
\end{equation*}
where $\mathbf{X}$ is a matrix of size $M\times N$, $\mathbf{W}$ is a matrix of size $N\times K$, and $\mathbf{Z}$ has size $M\times K$, as shown in Fig~\ref{fig:architecture}a.
This operation is performed using a semi-systolic array connected as an \ac{HWPE} to a PULP cluster (Fig.~\ref{fig:Pulp Cluster}).
Notice that, in DNN training, $\mathbf{X}$ and $\mathbf{W}$ can assume either input and weight matrices indifferently: the accelerator is designed symmetrically and can be indifferently used as weight- or input-stationary.

The accelerator is divided in several modules. First, the \textit{Datapath} is the core of RedMulE: it is an array of \ac{FP}16 \ac{FMA} units \cite{fpnew} interconnected in a semi-systolic fashion, as shown in Fig~\ref{fig:architecture}b. \ac{FMA} units are organized in \textit{L} rows, each made of \textit{H} columns. Within each row, a number of \textit{H} \acp{FMA} are wired together so that each \ac{FMA} computing an intermediate product will pass its result to the next one. The partial product computed by the last \ac{FMA} of a row is fed back as accumulation input of the first \ac{FMA} of the same row. Each \ac{FMA} features a design-time configurable number of internal pipeline registers (\textit{P}), and to saturate the array, the $\mathbf{X}$-matrix elements of each \ac{FMA} are held steady for $H\times(P+1)$ cycles, while new $W$-matrix operands are streamed-in cycle-by-cycle and broadcasted to all the \acp{FMA} of a column. Each row of \acp{FMA} computes $H\times(P+1)$ elements of a row of the $\mathbf{Z}$-matrix, that are stored in the memory only at the end of the computation, optimizing internal data reuse.

The \textit{Streamer} is a specialized memory access unit that connects RedMulE to the \ac{HCI} shallow branch through 9 \dbit{32} memory ports used alternatively for load and store operations. The Streamer is connected to three internal buffers: a X-Buffer that changes all the \textit{L} inputs of a column once every $H\times(P+1)$ cycles; a W-Buffer made of \textit{H} shift registers, each broadcasting a new $\mathbf{W}$-element to all the \textit{L} \acp{FMA} of a column every cycle; a Z-Buffer that buffers the computed $\mathbf{Z}$-elements.

The \textit{Scheduler} and the \textit{Controller} regulate the accelerator execution and contain the register file, accessed by the cores to program the accelerator.

We focus on a design with $H=4$, $L=8$, $P=3$ parameters, resulting in 32 \ac{FMA} units and 9 \dbit{32} \ac{TCDM} memory ports, for a \dbit{256} + \dbit{32} for non-word-aligned accesses. 

\begin{table*}[t]
\centering
\caption{State of the art comparison. First line = Best Efficiency; Second line = Peak Performance. 1 MAC = 2 OPs.}\label{table:comparison}
\resizebox{0.8\textwidth}{!}{
\begin{tabular}{|c|c|c|c|c|c|c|c|c|c|c|}
\hline
\multirow{2}{*}{\textbf{Category}}                                          & \multirow{2}{*}{\textbf{Design}} & \textbf{Tech} & \multicolumn{1}{l|}{\textbf{Area}} & \textbf{Freq}                                       & \textbf{Volt}                                       & \textbf{Power}                                       & \textbf{Perf}                                        & \textbf{Energy Eff}                                & \multirow{2}{*}{\textbf{\begin{tabular}[c]{@{}c@{}}MAC \\ Units\end{tabular}}} & \multirow{2}{*}{\textbf{Precision}} \\ 
                                                                            &                                  & \textit{nm}   & \multicolumn{1}{l|}{\textit{mm${}^{2}$}}  & \textit{MHz}                                        & \textit{V}                                          & \textit{mW}                                          & \multicolumn{1}{l|}{\textit{GOPS}}                   & \textit{GOPS/W}                                    &                                                                                &                                     \\ \hline
GPU                                                                         & NVIDIA A100 \cite{a100_data}     & 7             & -                                  & 1410                                                & -                                                   & 300000                                               & -                                                    & -                                                  & 256                                                                            & FP16                                \\ \hline
\multirow{5}{*}{\begin{tabular}[c]{@{}c@{}}Inference \\ Chips\end{tabular}} & Eyeriss \cite{eyeriss}           & 65            & 12.25                              & 250                                                 & 1.0                                                 & 278                                                  & 46                                                   & 166                                                & 168                                                                            & INT16                               \\ \cline{2-11} 
                                                                            & EIE \cite{eie}                   & 45            & 40.8                               & 800                                                 & -                                                   & 590                                                  & 102                                                  & 173                                                & 64                                                                             & INT8                                \\ \cline{2-11} 
                                                                            & Zeng etal. \cite{rssa}           & 65            & 2.14                               & 250                                                 & -                                                   & 478                                                  & 1152                                                 & 2410                                               & 256                                                                            & INT8                                \\ \cline{2-11} 
                                                                            & Simba \cite{simba}               & 16            & 6                                  & \begin{tabular}[c]{@{}c@{}}161\\ 2000\end{tabular}  & \begin{tabular}[c]{@{}c@{}}0.42\\ 1.2\end{tabular}  & \begin{tabular}[c]{@{}c@{}}-\\ -\end{tabular}        & \begin{tabular}[c]{@{}c@{}}-\\ 4000\end{tabular}     & \begin{tabular}[c]{@{}c@{}}9100\\ -\end{tabular}   & 1024                                                                           & INT8                                \\ \hline
\multirow{2}{*}{\begin{tabular}[c]{@{}c@{}}Training\\ Chips\end{tabular}}   & IBM \cite{ibm}                   & 7             & 19.6                               & \begin{tabular}[c]{@{}c@{}}1000\\ 1600\end{tabular} & \begin{tabular}[c]{@{}c@{}}0.55\\ 0.75\end{tabular} & \begin{tabular}[c]{@{}c@{}}4400\\ 13000\end{tabular} & \begin{tabular}[c]{@{}c@{}}8000\\ 12800\end{tabular} & \begin{tabular}[c]{@{}c@{}}1800\\ 980\end{tabular} & 4096                                                                           & FP16                                \\ \cline{2-11} 
                                                                            & Cambricon-Q \cite{cambricon}     & 45            & 888                                & 1000                                                & 0.6                                                 & 1030                                                  & 2000                                                 & 2240                                               & 1024                                                                           & INT8                                \\ \hline
HPC                                                                         & Manticore \cite{manticore}       & 22            & 888                                & \begin{tabular}[c]{@{}c@{}}500\\ 1000\end{tabular}  & \begin{tabular}[c]{@{}c@{}}0.6\\ 0.9\end{tabular}   & \begin{tabular}[c]{@{}c@{}}200\\ 900\end{tabular}    & \begin{tabular}[c]{@{}c@{}}25\\ 54\end{tabular}      & \begin{tabular}[c]{@{}c@{}}188\\ 50\end{tabular}   & 24                                                                             & FP64                                \\ \hline
\begin{tabular}[c]{@{}c@{}}Mat-Mul\\ Acc.\end{tabular}                      & Anders et al. \cite{anders}      & 14            & 0.024                              & \begin{tabular}[c]{@{}c@{}}2.1\\ 1090\end{tabular}  & \begin{tabular}[c]{@{}c@{}}0.26\\ 0.9\end{tabular}  & \begin{tabular}[c]{@{}c@{}}0.023\\ 82.7\end{tabular} & \begin{tabular}[c]{@{}c@{}}0.068\\ 34\end{tabular}   & \begin{tabular}[c]{@{}c@{}}2970\\ 420\end{tabular} & 16                                                                             & FP16                                \\ \hline
\multirow{2}{*}{\textbf{Our work}}                                                           & \textbf{PULP + RedMulE}       & 22            & 0.5                              & \begin{tabular}[c]{@{}c@{}}476\\ 666\end{tabular}   & \begin{tabular}[c]{@{}c@{}}0.65\\ 0.8\end{tabular}  & \begin{tabular}[c]{@{}c@{}}43.5\\ 90.7\end{tabular}  & \begin{tabular}[c]{@{}c@{}}30\\ 42\end{tabular}      & \begin{tabular}[c]{@{}c@{}}688\\ 462\end{tabular}  & 32                                                                             & FP16                                \\ \cline{2-11} 
                                                                            & \textbf{PULP + RedMulE}       & 65            & 3.85                               & 200                                                 & 1.2                                                 & 89.1                                                & 12.6                                                 & 152                                                & 32                                                                             & FP16                                \\ \hline
\end{tabular}}\end{table*}

\subsection{Working Principle}
RedMulE's operation starts by pre-loading the X-Buffer with \textit{L} rows from $\mathbf{X}$-matrix, each row made of 16 \ac{FP}16 elements ($\dbit{256}$ memory width$/\dbit{16}$ precision), namely $\mathbf{x _{0,0}}$ - $\mathbf{x_{0,15}}$ for Row\_0, $\mathbf{x_{1,0}}$ - $\mathbf{x_{1,15}}$ for Row\_1, and so on. Then, RedMulE loads a set of 16 $\mathbf{W}$-elements ($\mathbf{w_{0,0}}$ - $\mathbf{w_{0,15}}$) inside the first shift register of the W-buffer, broadcasting each of them to all the \textit{L} \acp{FMA} of the first column.

After 4 (=$P+1$) cycles, all the \textit{L} \acp{FMA} in the first column pass their computed partial products to all the \textit{L} \acp{FMA} of the second column. The accelerator loads another set of 16 $\mathbf{W}$-elements ($\mathbf{w_{1,0}}$ - $\mathbf{w_{1,15}}$) to broadcast them to all the \acp{FMA} in the second column. Once all the \textit{H} \acp{FMA} of a row have completed their computations, calculating a subset of 16 row-column intermediate products, RedMulE activates its feedback to provide the intermediate results to the accumulation input of the first \ac{FMA} of the given row and reiterates the computation. Then, the X-Buffer provides a new $\mathbf{X}$-operand to the first column of \acp{FMA}, and a new set of 16 $\mathbf{W}$-elements is re-loaded in the first W shift register. After four cycles, all the \textit{L} \acp{FMA} of the first column produce a new partial product and provide it to the \acp{FMA} in the second column. X-Buffer provides a new $\mathbf{X}$-operand at the input of the second column of \acp{FMA}, and the W-Buffer loads a new set of 16 $\mathbf{W}$-elements in the second W shift register for broadcasting, and the computation continues. Fig~\ref{fig:architecture}d shows the pipeline evolution inside a row of \acp{FMA}.

To guarantee a continuous data flow in the accelerator, the W-buffer accesses the memory once every 4-cycles to load a new set of 16 $\mathbf{W}$-elements. Once the X-Buffer is empty, ReMulE reuses the \textit{Streamer} port to load the $\mathbf{X}$-operands. Such operation is made by interleaving the memory accesses to $\mathbf{X}$-matrix between two adjacent $\mathbf{W}$-matrix accesses (Fig~\ref{fig:architecture}c) until the complete fulfillment of the X-buffer, maximizing the memory port utilization. After an entire row-column multiplication concludes, the Z-Buffer buffers the output products, and then also store operations are interleaved between two adjacent $\mathbf{W}$ load accesses until the Z-Buffer is empty.

\begin{figure*}[htbt]
\centerline{\includegraphics[width=\textwidth]{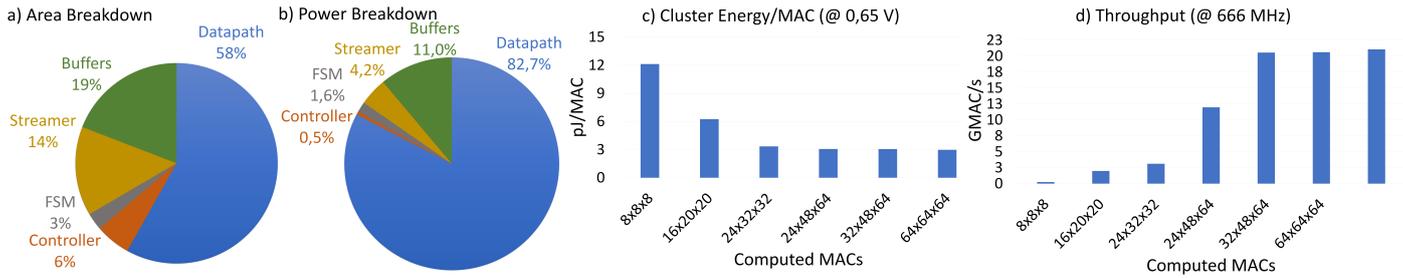}}
\caption{a) RedMulE area breakdown; b) RedMulE power breakdown; c) Cluster energy per \ac{MAC} operation; d) Throughput at maximum cluster frequency;\expandafter}
\label{fig:Figure3}
\end{figure*}

\section{Experimental Results} 
Our experiments target a \SI{22}{nm} technology using Synopsys Design Compiler for synthesis (slow corner at $f_\mathrm{targ}=\SI{208}{MHz}$, $V_{DD}=\SI{0.59}{V}$, $T=\SI{125}{\celsius}$) and Cadence Innovus for full-cluster Place\&Route in the same operating point. We performed timing and power analysis with back annotated switching activity from post-layout simulation in typical corner at $\SI{25}{\celsius}$, and \SI{0.65}{\volt} for peak energy efficiency or \SI{0.8}{V} for peak throughput and frequency. In Table~\ref{table:comparison}, we resume the State-of-the-Art comparison, including a \ac{PULP} cluster with RedMulE prototyped in \SI{65}{nm}.


\subsection{Area, efficiency and performance}
RedMulE occupies \SI{0.07}{mm^2}, corresponding to $14$\% of the entire PULP cluster. As can be noticed from Table~\ref{table:comparison}, at a system level, our design is the only one that occupies less than \SI{1}{mm^2}. The only exception is for Anders \etal~\cite{anders}, who show the results only for the systolic array alone. Fig.~\ref{fig:Figure3}a and Fig.~\ref{fig:Figure3}b show the area and power breakdown of the standalone RedMulE,
while Fig.~\ref{fig:Figure3}c shows the energy consumed by the cluster per FMA operation on RedMulE. At a cluster level, the average power consumption is \SI{43.5}{mW}, and the RedMulE contribution dominates it for $69$\%, while \ac{TCDM} banks and the \ac{HCI} interconnect contribution is $17.1$\%. Also in this case, our design outperforms the other accelerators. Only Anders \etal reach lower power consumption, but with a more scaled technology and in near-threshold operating conditions.

We reach a cluster peak energy efficiency of \SI{688}{GFLOPS/W}, which is $4.3\times$ lower than Anders \etal, whose results are obtained in near-threshold operating conditions and extremely reduced frequency (\SI{2.1}{MHz}). IBM~\cite{ibm} reaches $2.6\times$ better energy efficiency, at the cost of \SI{440}{mW} of power consumption, $10\times$ higher than our design.

\begin{figure*}[tb]
\centering
\includegraphics[width=\textwidth]{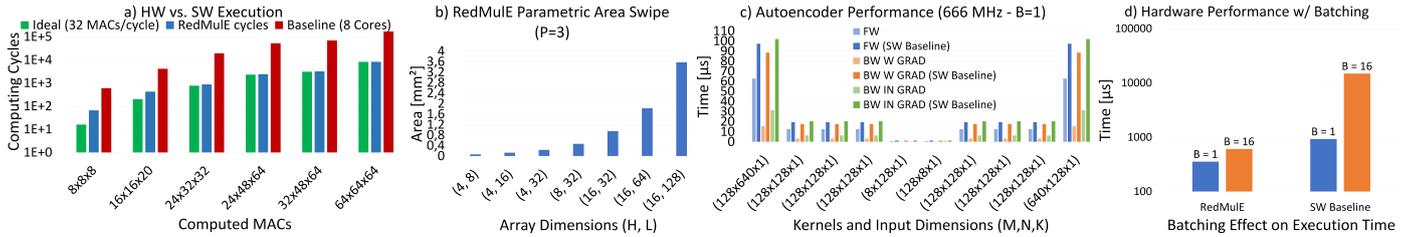}
\caption{a) HW vs. SW computational performancewith respect to ideal case (32 \acp{MAC}/cycle); b) RedMulE area swipe as a function of \textit{H} and \textit{L} ($P = 3$); c) RedMulE performance tested on Tiny\ac{ML} Perf Autoencoder benchmark; d) Effect of batching on HW/SW benchmark execution.}
\label{fig:Figure4}
\end{figure*}

Fig.~\ref{fig:Figure3}c and Fig.~\ref{fig:Figure3}d highlight how the RedMulE energy efficiency decreases when lowering the computational burden. For small matrix sizes, the control overhead proportionally increases, reducing the performance and thus the energy efficiency. It is evident that the cluster energy per \ac{FMA} operation considerably decreases when augmenting the amount of \ac{FMA} computation since the utilization increases.

We evaluated RedMulE's throughput and computation cycles against the SW execution on 8 parallel \riscv cores. RedMulE reaches a peak throughput of \SI{31.6}{MACs/cycle} ($98$\% utilization), meaning \SI{21.1}{GMACS}, or \SI{42}{GFLOPS} at \SI{666}{MHz} with \SI{0.80}{V} supply. Even though targeting different precision, our system performance is comparable to HPC designs. The Manticore prototype~\cite{manticore} features just $1.3\times$ higher performance despite its higher frequency, but with $10\times$ higher power consumption. On the other hand, we reach $1.2\times$ higher throughput than Anders \etal accelerator that works at $1.6\times$ higher frequency in the same precision.
To conclude, RedMulE introduces up to $22\times$ speedup over the software baseline and reaches $98.8$\% of the ideal case for a higher amount of computations (Fig.~\ref{fig:Figure4}a).

\subsubsection*{Parametric area swipe}

We studied the area overhead introduced when changing the number of \acp{FMA} within RedMulE, fixing the \ac{FMA}'s internal pipeline registers to $P=3$ (Fig~\ref{fig:Figure4}b).
RedMulE's area occupation becomes comparable to the area of the entire \ac{PULP} cluster with 256 \acp{FMA} ($H=8$, $L=32$), and doubles it with 512 ($H=16, L=32$).
%
Changing the shape of the internal array also affects the number of memory ports. In particular, changing the \textit{H} parameter from 4 to 5 results in including 4 ($=P+1$) additional pipeline registers within each row. To keep a high \acp{FMA} utilization, the bandwidth towards the memory increases by $4\times{\dbit{16}}$ (two additional memory ports), limiting the integration in the cluster.

\subsection{Use Case: TinyMLPerf AutoEncoder}
We evaluated RedMulE's performance on the Tiny\ac{ML}Perf \cite{tinymlperf} AutoEncoder benchmark as a possible use-case, compared with a software baseline executed on 8 \riscv cores. First, we conducted the test with a batch size (\textit{B}) of 1, i.e., a single input propagated forward and backward through the autoencoder (Fig~\ref{fig:Figure4}c).
RedMulE speedup is $2.6\times$ with significant advantages in particular in backward operations. The accelerator has a smaller speedup during forward operations due to the \textit{K} dimension, which is constant and equal to $B$. Consequently, RedMulE suffers from the effect of the latency introduced by the pipeline stages but does not benefit from the effect of the throughput because there are not sufficient elements in the activation matrix to fulfill the pipelines.
We can improve the utilization of the accelerator by increasing $B$, at the cost of more memory required for activations. We compare $B=1$ and $B=16$ in Fig~\ref{fig:Figure4}d; both configurations are well-fitting the L2 memory of a typical PULP-based system, with the $B=16$ one having an overall footprint of \SI{184}{kB}.
While the performance of the software baseline does not scale with a larger $B$, RedMulE's throughput is improved by almost $16\times$, achieving $24.4\times$ of speedup over the software counterpart.


\section{Conclusions}
We presented RedMulE, a cluster-coupled accelerator for \ac{FP}16 matrix multiplications that occupies \SI{0.07}{mm^2} ($14$\% of a \ac{PULP} cluster with 8 \riscv cores) and introduces up to $22\times$ speedup and $4.65\times$ higher energy efficiency than a software counterpart running on 8 \riscv cores.

\section*{Acknowledgement}
This work was supported in part by the EU H2020 "WiPLASH" (g.a. 863337), by the ECSEL H2020 "AI4DI" (g.a. 826060), and by Thales Alenia Space.


\bibliographystyle{IEEEtran}
\bibliography{biblio_abbrv}

\end{document}